\documentclass[preprint,aps,amsmath,amssymb,12pt,sort&compress,merge,prd]{revtex4-2}
\usepackage{graphicx}  % needed for figures
\usepackage{slashed}   % for Dirac Slash
\usepackage{subfig} % for sub fig
\usepackage{hyperref}

\begin{document}
%\widetext

\title{Detecting anomalous quartic gauge couplings using the isolation forest machine learning algorithm}
\author{Li Jiang}
\email{847495612@qq.com}
\author{Yu-Chen Guo}
\email{ycguo@lnnu.edu.cn}
\author{Ji-Chong Yang}
\email{yangjichong@lnnu.edu.cn}
\affiliation{Department of Physics, Liaoning Normal University, Dalian 116029, China}

\begin{abstract}
The search of new physics~(NP) beyond the Standard Model is one of the most important tasks of high energy physics.
A common characteristic of the NP signals is that they are usually few and kinematically different.
We use a model independent strategy to study the phenomenology of NP by directly picking out and studying the kinematically unusual events.
For this purpose, the isolation forest~(IF) algorithm is applied, which is found to be efficient in identifying the signal events of the anomalous quartic gauge couplings~(aQGCs).
The IF algorithm can also be used to constraint the coefficients of aQGCs.
As a machine learning algorithm, the IF algorithm shows a good prospect in the future studies of NP.
\end{abstract}

\maketitle

\section{\label{level1}Introduction}

Despite the great success of the Standard Model~(SM), there are still many unanswered questions making the search for new physics~(NP) beyond the SM a very important issue~\cite{johnellis}.
Except for a few cases that are known or suspected to deviate from the SM~\cite{neutrinomass1,*neutrinomass2,*g2muon,*p5prime,*rdstar1,*rdstar2}, in most cases, the experiments are in good agreement with the SM.
The searching of NP is to look for a small number of anomalies in the vast amount of experimental data.
Meanwhile, the kinematic features of the events induced by NP are usually different from the SM, which is the reason that event selection strategies~(ESSs) are commonly applied in the phenomenological studies of NP.
From the perspective of the SM effective theory~(SMEFT)~\cite{weinberg,*SMEFTReview1,*SMEFTReview2,*SMEFTReview3,*SMEFTReview4}, that is because the signals are induced by new interactions different from the SM.
It follows that the search of NP is to search the events which are `few and kinematically different'.
In this paper, we make use of the above features of NP and use a model independent strategy to directly search for kinematically unusual events.

This strategy has the advantage that it can be applied generally.
Taking the SMEFT for example, which is also a model independent way to search for the NP signals.
For one generation of fermions, there are $895$ baryon number conserving dimension-8 operators~\cite{vbscan,d8}, and kinematic analysis needs to be done for each operator.
Compared with the SMEFT, our strategy is independent of the operators.
Except for that, unlike the search of NP in a process which may turn out to be a wasted effort, the kinematically unusual events are always worth attention even if they are not from NP.
They could be faults in the experiments, or could be the rare processes allowed by the SM.

As an example, we use our strategy to study the anomalous quartic gauge couplings~(aQGCs)~\cite{aqgcold,*aqgcnew}, which are modifications to the SM gauge interactions intensively studied~\cite{aqgc1,*aqgc2,*aqgc4,*aqgc5,aqgc3,vbscan}.
aQGCs can be contributed by a lot of NP models such as Born-Infeld theory, composite Higgs, warped extra dimensions, two Higgs doublet models, $U(1)_{L_{\mu}-L_{\tau}}$, and axion-like particles~\cite{bi1,*bi2,composite1,*composite2,*extradim,*2hdm1,*2hdm2,*zprime1,*zprime2,*alp1,*alp2}.
Since dimension-6 operators cannot contribute to aQGCs independently~\cite{vbscan}, we concentrate on the dimension-8 operators.
A recent study shows that the dimension-8 operators are important in the convex geometry point of view to the SMEFT space~\cite{convexgeometry}.
Besides, there are cases sensitive to dimension-8 operators because the contributions from dimension-6 operators are absent~\cite{bi1,*bi2,ntgc1,ntgc2,*ntgc3,*ntgc4,*ntgc5}.
In the case of anomalous gauge couplings, aQGCs can lead to richer helicity combinations than dimension-6 anomalous trilinear gauge couplings~(aTGCs)~\cite{ssww}.
Besides, aQGCs can originate from tree diagrams while dimension-6 aTGCs are generated by loop diagrams~\cite{looportree}.
Consequently, while the SMEFT has mainly been applied with dimension-6 operators, the importance of dimension-8 operators has been pointed out in many previous studies~\cite{aqgcold,*aqgcnew,ntgc1,d8}.

The search of `few and kinematically different' events is in fact an anomaly detection~(AD), the applications of which in high energy physics~(HEP) are under developing extensively recently~\cite{ad}.
AD is suitable for machine learning~(ML), which has been used in various aspects of HEP~\cite{mlreview,*ml1,*ml2,*ml3,*ml4,*ml5,*ml6,*ml8,*ml10,*ml11}.
When it comes to AD, there are many algorithms such as autoencoder~\cite{autoencoder1,*autoencoder2,ml7,ml9}, multivariate Gaussian mixture model~\cite{guassian,ml9}, deep support vector data description~\cite{deeponeclass,ml7}, and isolation forest~(IF)~\cite{4781136,ml7,ml9}.
We use the IF algorithm because the mechanism behind IF algorithm is transparent, it merely identifies the points which are few and far away from the others.
Moreover, it is expected to perform better with fewer signal events, it is efficient to apply and easy to implement.
We find that the IF algorithm works as an automatic ESS and can identify the signal events very well.
Besides, IF algorithm can also be applied to constraint the parameters of NP models such as the coefficients of aQGCs, therefore has a lot of potential in the future studies of NP.

The rest of the paper is organized as follows, in Sec.~\ref{level2} we briefly introduce the IF algorithm; in Sec.~\ref{level3}, the application of IF algorithm on detecting the signals of aQGCs is presented; Sec.~\ref{level4} is a summary.

\section{\label{level2}A brief introduction of isolation forest}

IF algorithm is an algorithm with linear complexity designed for detection of point anomalies.
It makes use of the fact that the anomalies are `few and different'.
It can be applied for multi-dimensional data efficiently.
We briefly introduce the IF algorithm following Ref.~\cite{4781136}.

The key step of IF algorithm is to build an ensemble of isolation trees~(ITs).
The IT is a binary tree structure randomly generated to isolate every single point.
Denoting each point in the data set as $p^i(x_1^i,x_2^i,\ldots ,x_D^i)$, the construction procedure of an IT can be summarized as follows:
\begin{enumerate}
  \item Put all points into a root node.
  \item Randomly select a node which has not been partitioned yet.
  \item Randomly select a dimension $1\leq d\leq D$.
  \item Randomly set a split value ${\rm min}(x^i_d)< x<{\rm max}(x^i_d)$ where $i$ runs over all points in this node.
  \item Generate two children nodes, put the points with $x_d<x$ into the left child, and the others into the right child.
  \item Repeat (2) to (5) until every node is either partitioned or is filled with only one point.
\end{enumerate}

In this paper, we do not set a maximum depth for the ITs.
When an IT is generated, and the path length from a leaf to the root node can be used to determine whether the point represented by the leaf is an anomaly.
The path lengths of anomalies are generally shorter than those of normal points.

Because an IT is constructed randomly, it can be expected that the path lengths of points are not stable for a single IT.
Therefore, it will be more convincing to introduce multiple ITs, together as an IF.
Then the average path lengths over the ITs can be used to discriminate the anomalies from the normal points.

There are only two variables in this algorithm: the number of ITs, and the size of the data set.
As will be shown latter, the two variables can be made irrelevant of the problem.
More details and extensions of the IF algorithm can be found in Refs.~\cite{4781136,KARCZMAREK2020105659}.

\section{\label{level3}Application of IF algorithm on the search of aQGCs}

The IF algorithm can be applied in many different NP models.
In the absence of clear signs for NP, we use the detection of aQGCs as an example.

\subsection{\label{level3.1} aQGC signals in the process \texorpdfstring{$pp\to j j \ell ^+\ell ^-\nu\bar{\nu}$}{}}

The vector boson scattering~(VBS) processes at the LHC are very suitable for searching for the existence of aQGCs~\cite{vbscan,VBSReview1,*vbsadd1,*vbsadd2,*vbsadd3}.
They have been extensively studied by both ATLAS group and CMS group, and the effort will continue with future runs of the LHC.
After the first evidence of VBS process at the LHC found in 2014~\cite{sswwexp1}, a number of experimental results of VBS processes have been obtained~\cite{ssww,zaexp1,*zaexp2,*waexp1,*zzexp1,*zzexp2,*wzexp1,*wzexp2,*wwexp2,*coefficient1}.

Recently, the evidence of exclusive or quasi-exclusive $\gamma \gamma \to W^+W^-$ process has been found~\cite{wwexp1}.
As an illustration, we concentrate on this process at $\sqrt{s}=13\;{\rm TeV}$.
The next-to-leading order QCD corrections to the process $pp\to W^+W^- jj$ have been computed~\cite{vbfcut}, and the $K$ factor is found to be close to one ($K\approx 0.98$).
There are some difficulties in the phenomenological studies of NP in this process because of the presence of two neutrinos in $\ell^+\ell^-\nu \bar{\nu}jj$ which makes the reconstruction of the two $W$ bosons almost impossible.
However, these difficulties just provide a good test for the IF algorithm.

The Lagrangian relevant to this process is $\mathcal{L}_{\rm aQGC}=\sum _{i} (f_{M_i}/\Lambda^4)O_{M,i}+\sum _{j} (f_{T_j}/\Lambda^4)O_{T,j}$ with~\cite{aqgcold,*aqgcnew}
\begin{equation}
\begin{array}{ll}
O_{M,0}={\rm Tr\left[\widehat{W}_{\mu\nu}\widehat{W}^{\mu\nu}\right]}\times \left[\left(D^{\beta}\Phi \right) ^{\dagger} D^{\beta}\Phi\right],
 &O_{T,0}={\rm Tr}\left[\widehat{W}_{\mu\nu}\widehat{W}^{\mu\nu}\right]\times {\rm Tr}\left[\widehat{W}_{\alpha\beta}\widehat{W}^{\alpha\beta}\right],\\
O_{M,1}={\rm Tr\left[\widehat{W}_{\mu\nu}\widehat{W}^{\nu\beta}\right]}\times \left[\left(D^{\beta}\Phi \right) ^{\dagger} D^{\mu}\Phi\right],
 &O_{T,1}={\rm Tr}\left[\widehat{W}_{\alpha\nu}\widehat{W}^{\mu\beta}\right]\times {\rm Tr}\left[\widehat{W}_{\mu\beta}\widehat{W}^{\alpha\nu}\right],\\
O_{M,2}=\left[B_{\mu\nu}B^{\mu\nu}\right]\times \left[\left(D^{\beta}\Phi \right) ^{\dagger} D^{\beta}\Phi\right],
 &O_{T,2}={\rm Tr}\left[\widehat{W}_{\alpha\mu}\widehat{W}^{\mu\beta}\right]\times {\rm Tr}\left[\widehat{W}_{\beta\nu}\widehat{W}^{\nu\alpha}\right],\\
O_{M,3}=\left[B_{\mu\nu}B^{\nu\beta}\right]\times \left[\left(D^{\beta}\Phi \right) ^{\dagger} D^{\mu}\Phi\right],
 &O_{T,5}={\rm Tr}\left[\widehat{W}_{\mu\nu}\widehat{W}^{\mu\nu}\right]\times B_{\alpha\beta}B^{\alpha\beta},\\
O_{M,4}=\left[\left(D_{\mu}\Phi \right)^{\dagger}\widehat{W}_{\beta\nu} D^{\mu}\Phi\right]\times B^{\beta\nu},
 &O_{T,6}={\rm Tr}\left[\widehat{W}_{\alpha\nu}\widehat{W}^{\mu\beta}\right]\times B_{\mu\beta}B^{\alpha\nu},\\
O_{M,5}=\left[\left(D_{\mu}\Phi \right)^{\dagger}\widehat{W}_{\beta\nu} D_{\nu}\Phi\right]\times B^{\beta\mu} + h.c.,
 &O_{T,7}={\rm Tr}\left[\widehat{W}_{\alpha\mu}\widehat{W}^{\mu\beta}\right]\times B_{\beta\nu}B^{\nu\alpha},\\
O_{M,7}=\left(D_{\mu}\Phi \right)^{\dagger}\widehat{W}_{\beta\nu}\widehat{W}_{\beta\mu} D_{\nu}\Phi, & \\
\end{array}
\label{eq.2.3}
\end{equation}

The subprocess $\gamma\gamma \to W^+W^-$ can be affected by the aQGCs via five vertices, they are $\mathcal{L}_{\gamma\gamma WW}=\sum _{i=0}^4 \alpha_i V_{\gamma\gamma WW_i}$ with
\begin{equation}
\begin{array}{ll}
V_{0}=F_{\mu\nu}F^{\mu\nu}W^{+\alpha}W^-_{\alpha}, &V_{1}=F_{\mu\nu}F^{\mu\alpha}W^{+\nu}W^-_{\alpha},\\
V_{2}=F_{\mu\nu}F^{\mu\nu}W^+_{\alpha\beta}W^{-\alpha\beta}, &V_{3}=F_{\mu\nu}F^{\nu\alpha}W^+_{\alpha\beta}W^{-\beta\mu},\\
V_{4}=F_{\mu\nu}F^{\alpha\beta}W^+_{\mu\nu}W^{-\alpha\beta},& \\
\end{array}
\label{eq.2.6}
\end{equation}
where $W^{\pm\mu\nu}\equiv \partial _{\mu}W^{\pm}_{\nu}-\partial _{\nu}W^{\pm}_{\mu}$.
The corresponding coefficients of vertices are
\begin{equation}
\begin{array}{ll}
\alpha_0=\frac{e^2v^2}{8\Lambda ^4}\left(f_{M_0}+\frac{c_W}{s_W}f_{M_4}+2\frac{c_W^2}{s_W^2}f_{M_2}\right),
&\alpha_1=\frac{e^2v^2}{8\Lambda ^4}\left(\frac{1}{2}f_{M_7}+2\frac{c_W}{s_W}f_{M_5}-f_{M_1}-2\frac{c_W^2}{s_W^2}f_{M_3}\right),\\
\alpha_2=\frac{1}{\Lambda ^4}\left(s_W^2f_{T_0}+c_W^2f_{T_5}\right),\;
&\alpha_3=\frac{1}{\Lambda ^4}\left(s_W^2f_{T_2}+c_W^2f_{T_7}\right),\;\\
\alpha_4=\frac{1}{\Lambda ^4}\left(s_W^2f_{T_1}+c_W^2f_{T_6}\right). &
\end{array}
\label{eq.2.7}
\end{equation}
Because each dimension-8 operator contributes to only one vertex, and because the constraints on dimension-8 operators are obtained by assuming one operator at a time in experiments, the constraints on $\alpha _i$ can be derived by the constraints on dimension-8 operators~\cite{aqgc3} and are listed in Table~\ref{tab.1}.

\begin{table}[!htbp]
\centering
\begin{tabular}{cc|cc}
\hline
vertex & constraint & coefficient & constraint\\
\hline
$\alpha_0 ({\rm TeV^{-2}})$ & $[-0.013, 0.013]$ & $f_{M_2}/\Lambda ^4\;({\rm TeV^{-4}})$ & $[-2.8, 2.8]$~\cite{waconstraint} \\
$\alpha_1 ({\rm TeV^{-2}})$ & $[-0.021, 0.021]$ & $f_{M_5}/\Lambda ^4\;({\rm TeV^{-4}})$ & $[-8.3, 8.3]$~\cite{waconstraint} \\
$\alpha_2 ({\rm TeV^{-4}})$ & $[-0.38, 0.38]$ & $f_{T_5}/\Lambda ^4\;({\rm TeV^{-4}})$ & $[-0.5, 0.5]$~\cite{waconstraint} \\
$\alpha_3 ({\rm TeV^{-4}})$ & $[-1.47, 1.69]$ & $f_{T_7}/\Lambda ^4\;({\rm TeV^{-4}})$ & $[-1.91, 2.12]$~\cite{zaexp3} \\
$\alpha_4 ({\rm TeV^{-4}})$ & $[-0.95, 0.97]$ & $f_{T_6}/\Lambda ^4\;({\rm TeV^{-4}})$ & $[-1.23, 1.26]$~\cite{zaexp3} \\
\hline
\end{tabular}
\caption{\label{tab.1}The constraints on vertices and the corresponding limits on the dimension-8 operators at 95\% CL.
}
\end{table}

\subsection{\label{level3.2}Detection of the signals}

%\begin{table}
%\begin{center}
%\begin{tabular}{c|c|c|c|c|c}
%\hline
%  & $V_0$ & $V_1$ & $V_2$ & $V_3$ & $V_4$ \\
%\hline
%$N_{\rm SM}$ & $16846$ & $95813$ & $58962$ & $37390$ & $5.11\times 10^6$  \\
%$N_{t\bar{t}}$ & $23654$ & $134531$ & $82789$ & $52500$ & $7.175\times 10^6$ \\
%\end{tabular}
%\end{center}
%\caption{\label{Tab:cutflow}The event number of SM backgrounds and $t\bar{t}$ background when $N_{\rm aQGCs}=50$.}
%\end{table}

\begin{figure}[!htbp]
\centering{
\includegraphics[width=0.7\textwidth]{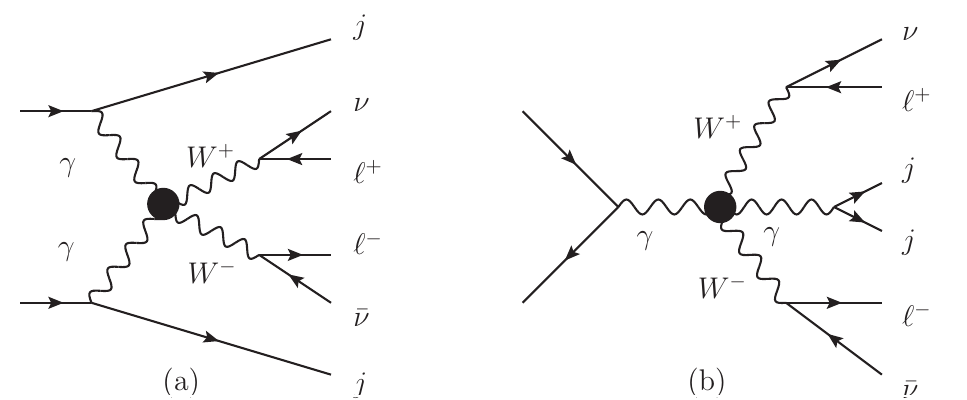}
\caption{\label{fig:feyndiag1}Typical Feynman diagrams for the signal.}}
\end{figure}

\begin{figure}[!htbp]
\centering{
\includegraphics[width=0.9\textwidth]{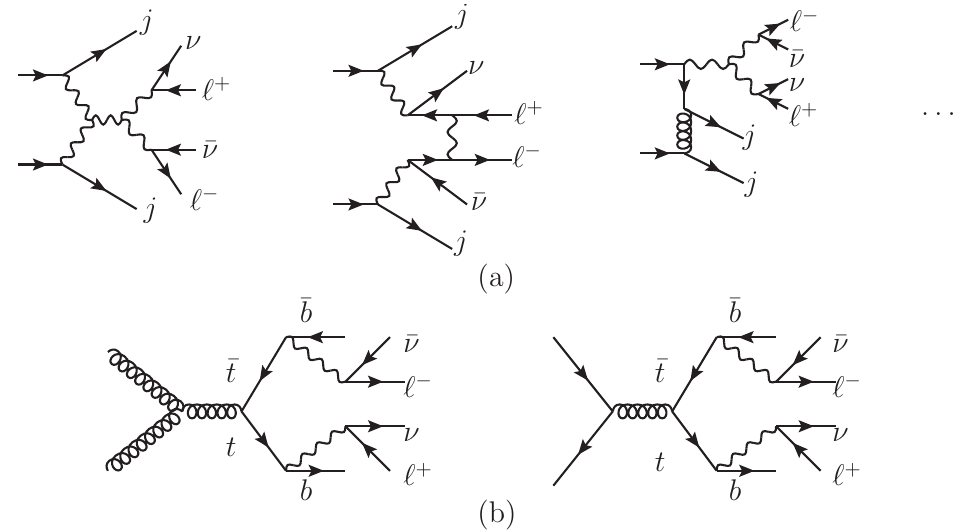}
\caption{\label{fig:feyndiag2}Typical Feynman diagrams for backgrounds.}}
\end{figure}

In this subsection, we assume the existence of the aQGCs and investigate whether the signal events can be picked out by IF algorithm.
The dominant signal is $W^+W^-jj$ production induced by aQGCs with leptonic decays of $W^{\pm}$ bosons as shown in Fig.~\ref{fig:feyndiag1}.~(a).
This process can also be contributed by the triboson production induced by aQGCs shown in Fig.~\ref{fig:feyndiag1}.~(b).
The background is the process $pp\to jj\ell^+\ell^- \nu\bar{\nu}$ in the SM, the typical diagrams are shown in Fig.~\ref{fig:feyndiag2}.~(a).
Except for that, we also consider the $t\bar{t}$ production with b-jet mistagged as dipicted in Fig.~\ref{fig:feyndiag2}.~(b), the b-tag efficiency is assumed to by $77\%$~\cite{btag}.
For simplicity, we neglect the triboson channel induced by aQGCs and the interference between the contributions from aQGCs and the SM which were found to be negligible~\cite{aqgc3}.
In the following we consider one operator at a time, therefore the interferences between different aQGCs are also neglected.

\begin{table}[!htbp]
\begin{center}
\begin{tabular}{c|c|c|c|c|c|c|c}
\hline
 &$\alpha _0=0.013$ & $\alpha _1=0.021 $ & $\alpha _2=0.38 $ & $\alpha _3=1.69$ & $\alpha _4=0.97$& SM & $t\bar{t}$ \\
&(TeV$^{-2}$)&(TeV$^{-2}$)&(TeV$^{-4}$)&(TeV$^{-4}$)&(TeV$^{-4}$)& \\
\hline
$\sigma$($\rm fb$) after $N_{\ell,j}$ cut                   &$0.91$   &$0.16$  &$0.26$  &$0.41$& $0.20$  & $306.6$ & $430.5$ \\
% $M_{jj}>150{\rm\ GeV},\ \Delta y_{jj}>1.2$     &$66.9$   &$12.6$  &$13.1$  &$4.0$& $0.91$ & $163.4$ & $3813$ \\
% $|\cos (\phi _{LM})| > 0.3$                    &$60.9$   &$11.7$  &$12.4$  &$3.7$& $0.84$ & $134.2$ & $3496$ \\
% $\cos(\theta _{ll})< 0$                        &$58.9$   &$11.3$  &$12.0$  &$3.6$& $0.80$ & $49.3$  & $1639$\\
% $\hat{s} > 1.5 \;{\rm TeV}^2$                  &$56.1$   &$10.8$  &$11.6$  &$3.4$& $0.75$ & $4.9$   & $151.4$ \\
\hline
\end{tabular}
\end{center}
\caption{\label{tab.cuts} The cross sections after $N_{\ell,j}$ cut.}
\end{table}

The events are generated by using Monte Carlo~(MC) simulation with \verb"MadGraph5_aMC@NLO"~\cite{madgraph,*feynrules}, including a parton shower with \verb"Pythia82"~\cite{pythia} and a CMS-like detector simulation with \verb"Delphes"~\cite{delphes}.
The basic cuts are set as same as the default settings of \verb"MadGraph5_aMC@NLO".
The parton distribution function is \verb"NNPDF2.3"~\cite{NNPDF}. The data preparation is performed by \verb"MLAnalysis"~\cite{Guo:2023nfu}.

To ensure the reliability, we require the particles in the final states to satisfy $N_{\ell^{\pm}}\geq 1,2\leq N_j \leq 5$ where $N_{\ell^{\pm}}$ are the numbers of (anti)leptons, $N_j$ is the number of jets.
This requirement is denoted as $N_{\ell,j}$ cut.
The cross sections of the signals and backgrounds after this cut are listed in Table~\ref{tab.cuts}.
For illustration, we concentrate on $V_{0,3}$ vertices which originate from $O_{M_i}$ and $O_{T_i}$ operators, respectively.
Denoting $N_{\rm SM,t\bar{t},aQGC}$ as event numbers of the SM background, $t\bar{t}$ background and the signal, we generate the events in the ratio $N_{\rm SM}:N_{t\bar{t}}:N_{\rm aQGC}=\sigma _{\rm SM}:\sigma _{t\bar{t}}:\sigma _{\rm aQGC}$, where $\sigma _{\rm SM,t\bar{t},aQGC}$ are cross sections of the SM background, $t\bar{t}$ background and the signal, respectively.
For the signals, we keep $N_{\rm aQGC}=50$ after the $N_{\ell,j}$ cut, and therefore the data sets are consist of events with $N_{\rm SM}:N_{t\bar{t}}:N_{\rm V_0}=16846:23654:50$ and $N_{\rm SM}:N_{t\bar{t}}:N_{\rm V_3}=37390:52500:50$.
Each event in the data set is assembled straightforwardly and is consist of $18$ attributes, which are components of transverse missing momentum $\slashed{\bf p}_T$, the 4-momenta of the hardest two jets $p_{j_1}$ and $p_{j_2}$, and the 4-momenta of the hardest (anti)lepton $p_{\ell^+}$ and $p_{\ell^-}$.

\begin{figure}[!htbp]
\centering{
\includegraphics[width=0.75\textwidth]{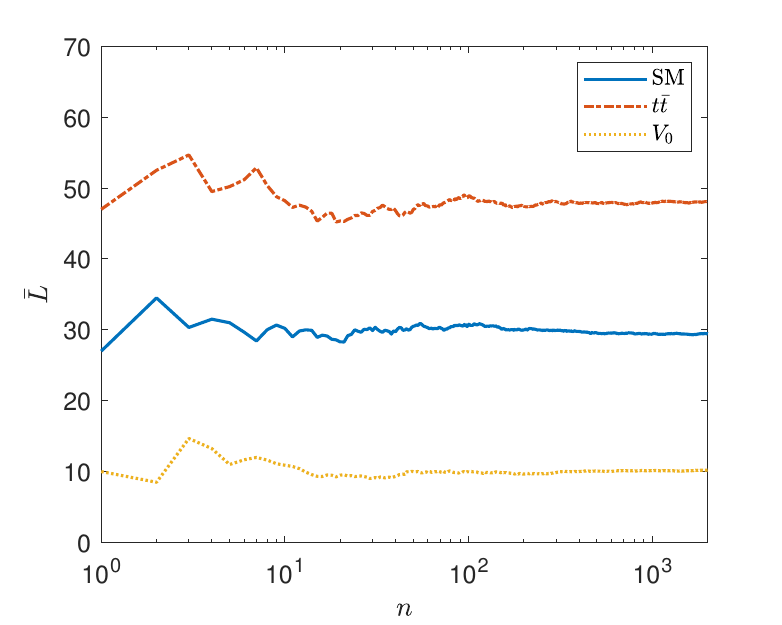}
\caption{\label{fig:converge}$\bar{L}$ as a function of $n$ for $V_0$ data set. Three events from different sources are picked randomly as examples. $\bar{L}$ is the average path length, $n$ is the number of isolation trees.}}
\end{figure}

There are two parameters in the IF algorithm.
One of the parameters is the number of trees, which is denoted as $n$.
$n$ is a model-independent parameter used to control the accuracy of the IF algorithm.
Denoting the path lengths as $L$, we find that $L$ converges quickly with growing $n$.
Picking one event out of each of the SM background, $t\bar{t}$ background and $V_0$ signal, as shown in Fig.~\ref{fig:converge}, $\bar{L}$ becomes stable after constructing about $1000$ trees.
In this paper, we use $n=2000$, the relative standard errors of $L$ are about $1\%$~($0.4\%-1.4\%$) for each point.

The other parameter is the size of the data set.
An anomaly score~(denoted as $a$) which is independent of the size of the data set can be defined by normalizing the average path length~(denoted as $\bar{L}$) with the average depth of an isolation tree $c(N)$ as $a=2^{-\bar{L}/c(N)}$, where $N$ is the size of the data set.~\cite{4781136},
\begin{equation}
c(N)=2H(N-1)-2(N-1)/N,
\label{eq.3.2.1}
\end{equation}
where $H(N)$ is the harmonic number.
$a$ is bounded in $(0,1)$.
When $a$ is larger, the corresponding event is more suspicious of anomalies.

\begin{figure}[!htbp]
\centering{
\includegraphics[width=0.48\textwidth]{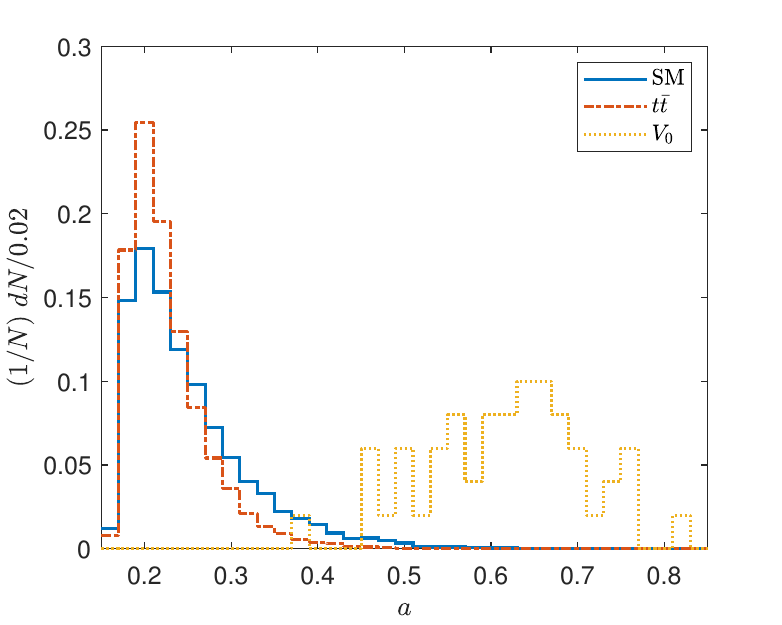}
\includegraphics[width=0.48\textwidth]{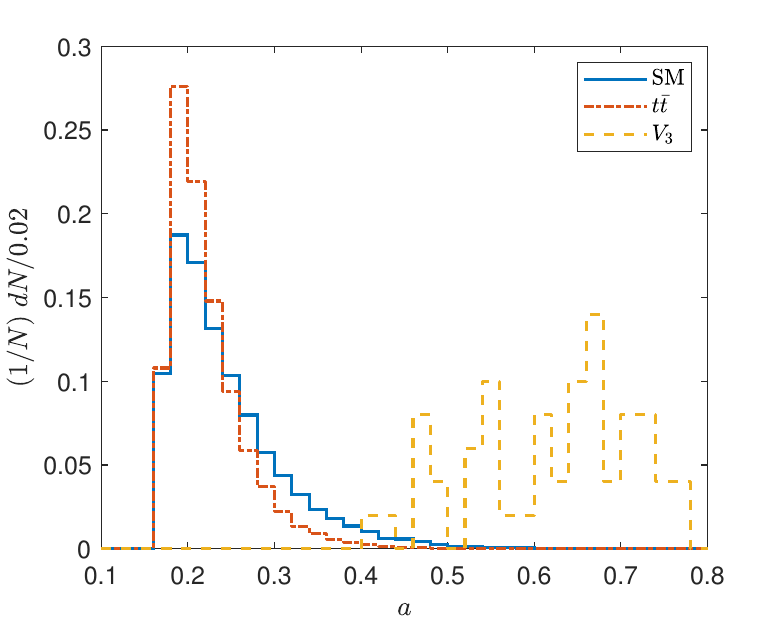}
\caption{\label{fig:dist}Normalized distributions of $a$, the left panel is for $V_0$ and the right panel is for $V_3$.}}
\end{figure}

\begin{figure}[!htbp]
\centering{
\includegraphics[width=0.48\textwidth]{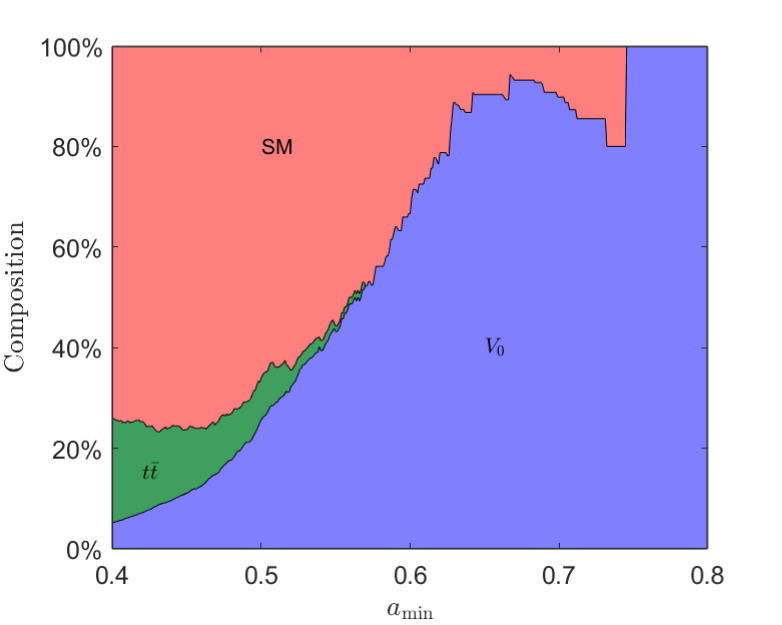}
\includegraphics[width=0.48\textwidth]{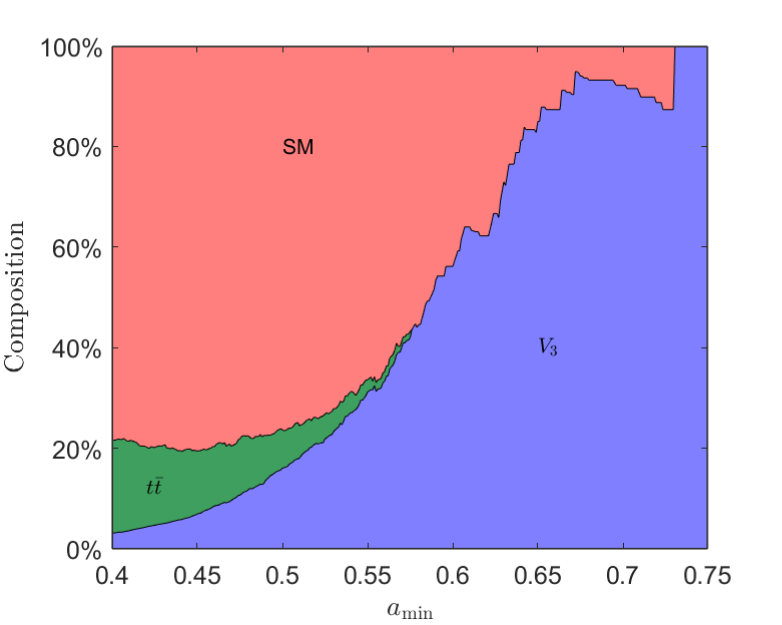}
\caption{\label{fig:comp}Compositions of the selected events passing a cut on anomaly score $a> a_{\rm min}$, the left panel is for $V_0$ and the right panel is for $V_3$.}}
\end{figure}

The normalized distributions of $a$ are shown in Fig.~\ref{fig:dist}.
We find that in both cases of $V_0$ and $V_3$, $a$ for the backgrounds are very different from those for the signals.
One can set a minimal anomaly score, and use $a>a_{\rm min}$ to pick out the signal events of aQGCs.
The compositions of the selected events are shown in Fig.~\ref{fig:comp}.
For both cases, with $a_{\rm min}=0.6$, about half of the selected events are signal events.
We find that the IF algorithm is powerful to pick out the signal events without the knowledge of the NP as long as the signal exists.

\subsection{\label{level3.3}Use the IF algorithm as an event selection strategy}

\begin{figure}[!htbp]
\centering{
\includegraphics[width=0.75\textwidth]{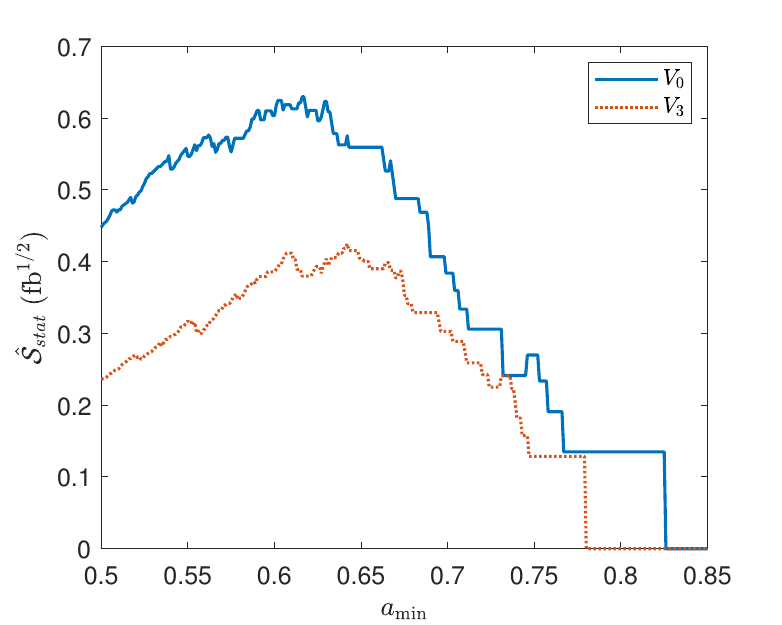}
\caption{\label{fig:cs}$\hat{\mathcal{S}}_{stat}=\sigma_s/\sqrt{\sigma_{bg}+\sigma_s}$ as functions of $a_{\rm min}$, where $\sigma _{s,bg}$ are the cross-sections of aQGC contribution and backgrounds after a cut on anomaly score $a>a_{\rm min}$.}}
\end{figure}
The effect of the IF algorithm is similar to an event selection strategy~(ESS).
Different from the traditional ESS, for the IF algorithm there is no need to study the kinematic features.
IF algorithm is like an automatic ESS which can be generally applied for a large class of NP signals.

In the search of NP, the signal significance is widely used which is defined as $\mathcal{S}_{stat}=N_s/\sqrt{N_{bg}+N_s}$ where $N_{s,bg}$ are event numbers of signal and background.
Similarly, a luminosity independent quantity can be defined as $\hat{\mathcal{S}}_{stat}=\sigma_s/\sqrt{\sigma_{bg}+\sigma_s}$ such that $\mathcal{S}_{stat}=\sqrt{l} \hat{\mathcal{S}}_{stat}$ where $l$ is the luminosity.
In this paper, we use $\hat{\mathcal{S}}_{stat}$ to qualify the ESS.
By selecting events with $a>a_{\rm min}$, $\hat{\mathcal{S}}_{stat}$ for $V_{0,3}$ are shown in Fig.~\ref{fig:cs}.
The $\hat{\mathcal{S}}_{stat}$ can reach $0.630\;{\rm fb}^{1/2}$ at $a_{\rm min}=0.617$ for $V_0$ and $0.425\;{\rm fb}^{1/2}$ at $a_{\rm min}=0.642$ for $V_3$.

We compare the IF algorithm with the ESS designed for the aQGCs in the process $pp\to jj\ell^+\ell^-\nu\bar{\nu}$ proposed in Ref.~\cite{aqgc3}, which are
\begin{equation}
\begin{split}
&M_{jj}>150{\rm\ GeV},\ \Delta y_{jj}>1.2,\;\;|\cos (\phi _{LM})| > 0.3,\\
&\cos(\theta _{\ell\ell})< 0,\;\;\hat{s} > 1.5 \;{\rm TeV}^2,\;\;M_{o1}>600{\rm\ GeV},
\end{split}
\label{eq.3.2.2}
\end{equation}
where $M_{jj}$ and $\Delta y_{jj}$ are invariant mass and difference between the rapidities of the hardest two jets, $\phi _{LM}$ is the angle between sum of the transverse momenta of charged leptons ${\bf p}^{\ell^+}_T+{\bf p}^{\ell^-}_T$ and $\slashed {\bf p}_T$, $\theta _{\ell\ell}$ is the angle between the charged leptons, and
\begin{equation}
\begin{split}
&M_{o1}\equiv \sqrt{\left(|{\bf p}_T^{\ell^+}|+|{\bf p}_T^{\ell^-}|+|\slashed{\bf p}_{\rm T}|\right)^2-\left|{\bf p}_T^{\ell^+}+{\bf p}_T^{\ell^-}+\slashed{\bf p}_{\rm T}\right|^2},\\
&\hat{s}=\left((1+|u|)E^{\ell^+}+(1+|v|)E^{\ell^-}\right)^2-\left((1+u){\bf p}_z^{\ell^+}+(1+v){\bf p}_z^{\ell^-}\right)^2 - \left|\sum _{\pm}\mathbf{p}^{\ell^\pm}_T+\slashed{\bf p}_T\right|^2,
\end{split}
\label{eq.3.2.3}
\end{equation}
with $E^{\ell^{\pm}}$ the energies of charged leptons, and
\begin{equation}
\begin{split}
&u=\frac{1}{\kappa}\left(\slashed {\bf p}_y {\bf p}^{\ell^-}_x -  \slashed{\bf p}_x {\bf p}^{\ell^-}_y\right), \;\;
  v = -\frac{1}{\kappa}\left(\slashed {\bf p}_y {\bf p}^{\ell^+}_x - \slashed {\bf p}_x {\bf p}^{\ell^+}_y\right),\;\;
  \kappa = {\bf p}^{\ell^+}_y {\bf p}^{\ell^-}_x - {\bf p}^{\ell^+}_x {\bf p}^{\ell^-}_y.
\end{split}
\label{eq.3.2.4}
\end{equation}

\begin{figure}[!htbp]
\centering{
\includegraphics[width=0.48\textwidth]{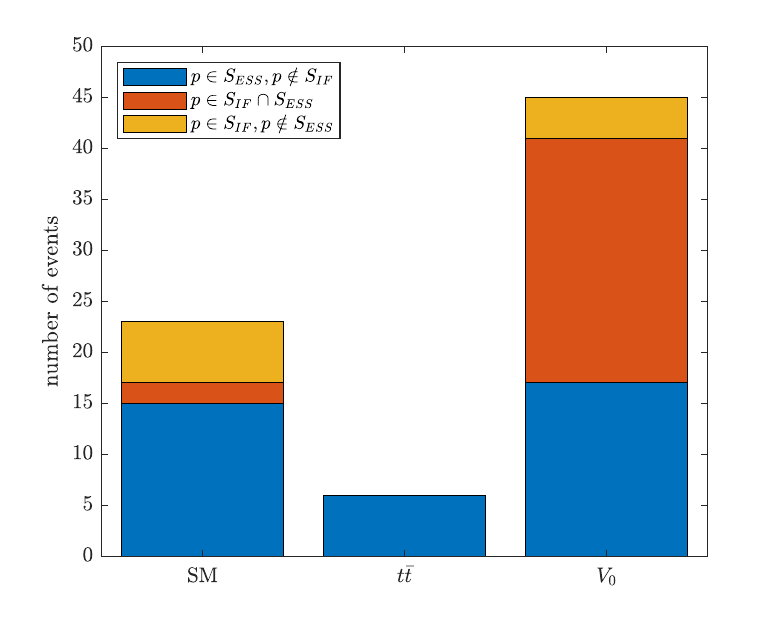}
\includegraphics[width=0.48\textwidth]{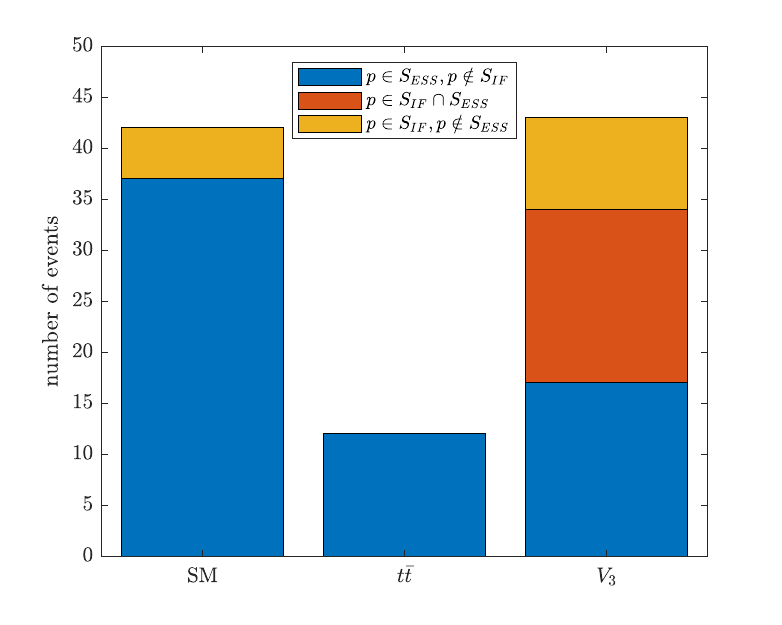}
\caption{\label{fig:compare}Difference between the selected events by using Eq.~(\ref{eq.3.2.2}) and using anomaly scores. The blue bars show the number of events selected by Eq.~(\ref{eq.3.2.2}) but not by anomaly score cuts, the yellow bars show the number of events selected by anomaly score cuts but not by Eq.~(\ref{eq.3.2.2}), the orange bars show the number of events selected by both methods.}}
\end{figure}
For IF algorithm, we select events with $a>0.617$ and $a>0.642$ for $V_0$ and $V_3$, respectively, the result sets are denoted as $S_{IF}$.
The sets consisting of events selected by Eq.~(\ref{eq.3.2.2}) are denoted as $S_{ESS}$.
The numbers of events in those sets are shown in Fig.~\ref{fig:compare}.
As one can see that the events picked by IF algorithm is not quite the same as the ESS in Eq.~(\ref{eq.3.2.2}), especially for the backgrounds.
The $\hat{\mathcal{S}}_{stat}$ for $V_{0,3}$ with Eq.~(\ref{eq.3.2.2}) are $0.691\;{\rm fb}^{1/2}$ and $0.341\;{\rm fb}^{1/2}$.
Compared with the results of IF algorithm, we find that the ESS using anomaly score shows competitive ability in discriminating signals, especially for the cases that the signal events are fewer.

\begin{figure}[!htbp]
\centering{
\includegraphics[width=0.7\textwidth]{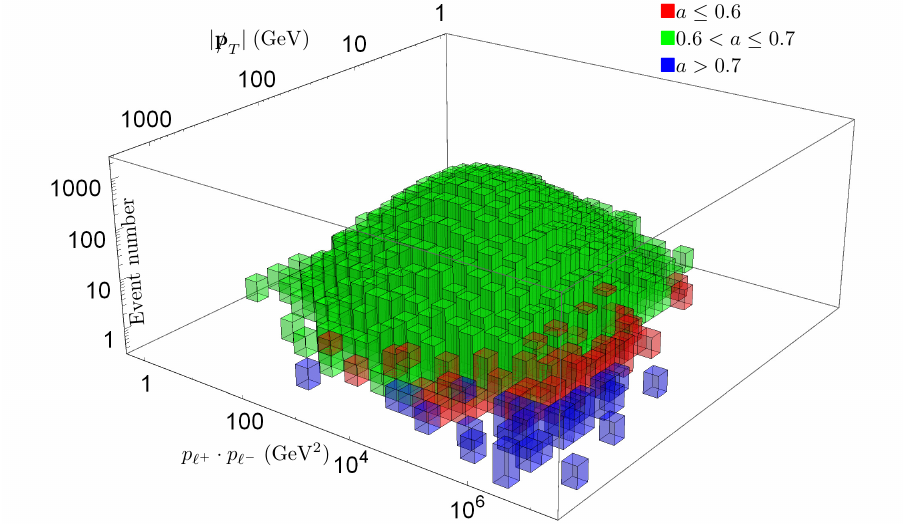}\\
\includegraphics[width=0.7\textwidth]{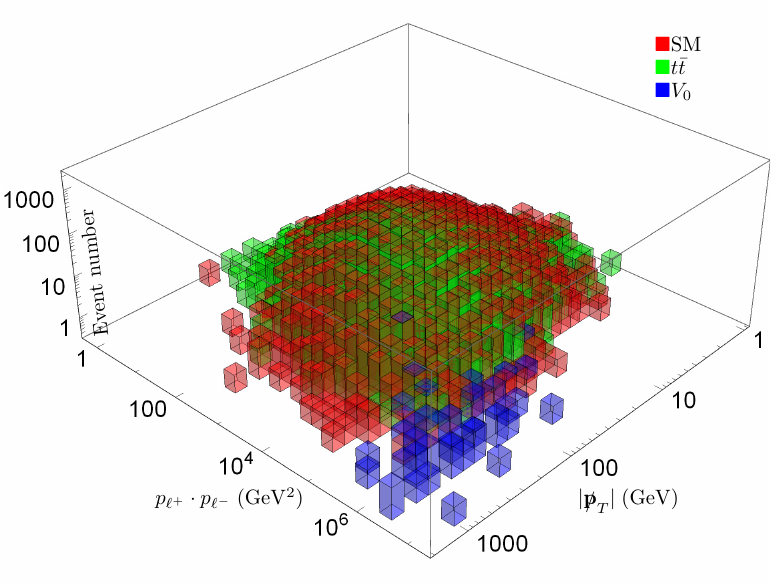}
\caption{\label{fig:dist3d}Distributions of events in the $(|\slashed{\bf p}_T|, p_{\ell^+}\cdot p_{\ell ^-})$ space. There $31\times 31$ bins. The widthes are $[\exp(0.251 (k-1)) , \exp(0.251 k))$ (GeV) for $|\slashed{\bf p}_T|$ dimension and $[\exp(0.558 (k-1)) , \exp(0.558 k))$ $({\rm GeV}^2)$ for $|\slashed{\bf p}_T|$ dimension where $k$ are integers with $0\leq k \leq 30$.}}
\end{figure}

In the above we using $18$ attributes which is straightforward, but not optimized.
There are usually observables more sensitive to the signal, which depend on the model or operators one looks for.
For example, knowing that we are searching for aQGCs, we can use attributes such as $M_{o1}$, $|\slashed{\bf p}_T|$ and $p_{\ell^+}\cdot p_{\ell ^-}$.
By choosing only two attributes, $|\slashed{\bf p}_T|$ and $p_{\ell^+}\cdot p_{\ell ^-}$, the events can be represented by points in a $2D$ space, and therefore easy to visualize.
By applying IF algorithm on these attributes, the distributions of events with different anomaly scores are shown in Fig.~\ref{fig:dist3d}, and one can see that the events with the higher anomaly scores are indeed those events far away from the others.
The distributions of the events from backgrounds and signal are also shown in Fig.~\ref{fig:dist3d}, which indicate that the events far away from the others are indeed the signal events.

\subsection{\label{level3.4}Set constraints on the coefficients}

A more common scenario is that signal events are not observed and one needs to set constraints on the parameters of NP models and coefficients of operators.
This can also been done with the help of the IF algorithm, because the mechanism behind the IF algorithm suggests that the anomaly scores of the backgrounds should not be sensitive to the signal events.
Consequently, after constructing an IF for the MC data of the backgrounds, which is model independent, one can use anomaly score as a cut, the expected cross section after this cut can be calculated, and can be compared with the cross section obtained by experiments under the same cut.
However, when it comes to constraint the parameters of a specific model, we need the information of this model which is not model independent any more.

Again, we take $V_{0,3}$ vertices as examples to illustrate this approach.
To set constraints on the coefficients, the data sets are assembled with $N_{\rm SM}:N_{t\bar{t}}:N_{\rm aQGC}$ with $N_{\rm SM},N_{t\bar{t}}$ as same as the previous section, and with $N_{\rm aQGC}=0,10,20,30,40,50$.
They correspond to $\alpha _0 = \sqrt{m/5}\times 0.013$ ${\rm TeV^{-2}}$ and $\alpha _3 = \sqrt{m/5}\times 1.69$ ${\rm TeV^{-4}}$ with $m=0,1,\ldots,5$ when neglecting the interferences.

\begin{figure}[!htbp]
\centering{
\includegraphics[width=0.48\textwidth]{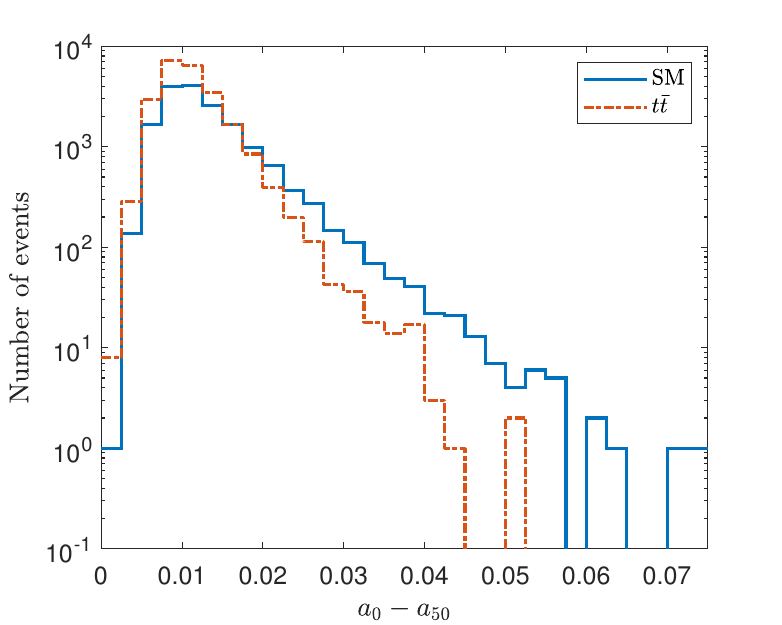}
\includegraphics[width=0.48\textwidth]{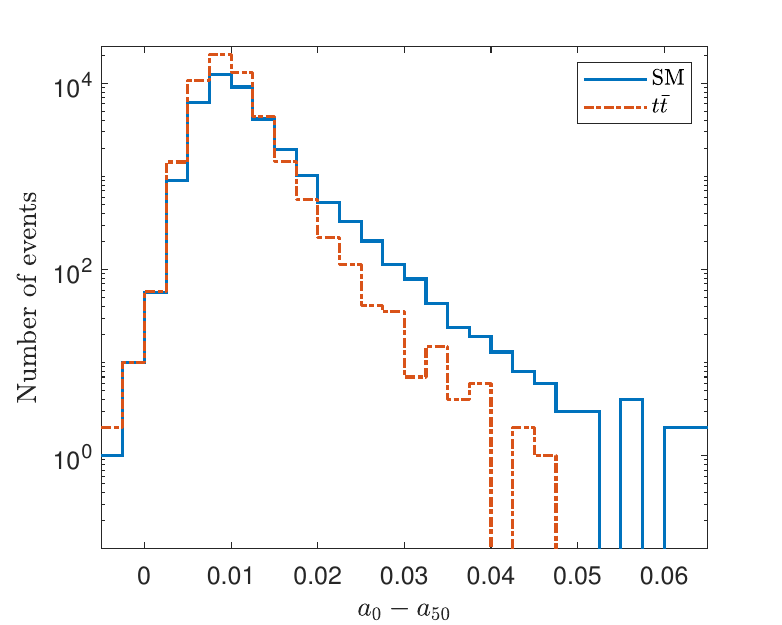}
\caption{\label{fig:aschange}The distributions of $a_0-a_{50}$ for the backgrounds, where $a_0-a_{50}$ are the changes of the anomaly scores from a data set without signal events to a data set with $50$ signal events.}}
\end{figure}

Denoting $a_{0,50}$ as anomaly scores of events for the $N_{\rm aQGC}=0$ and $N_{\rm aQGC}=50$ data sets, respectively.
The distributions of $a_0-a_{50}$ for the backgrounds are shown in Fig.~\ref{fig:aschange}.
We find that, the anomaly scores of the backgrounds increase a little bit without the signal events.
For $V_0$, $0<a_0-a_{50}<0.075$ and for $V_3$, $-0.005<a_0-a_{50}<0.065$.

\begin{figure}[!htbp]
\centering{
\includegraphics[width=0.48\textwidth]{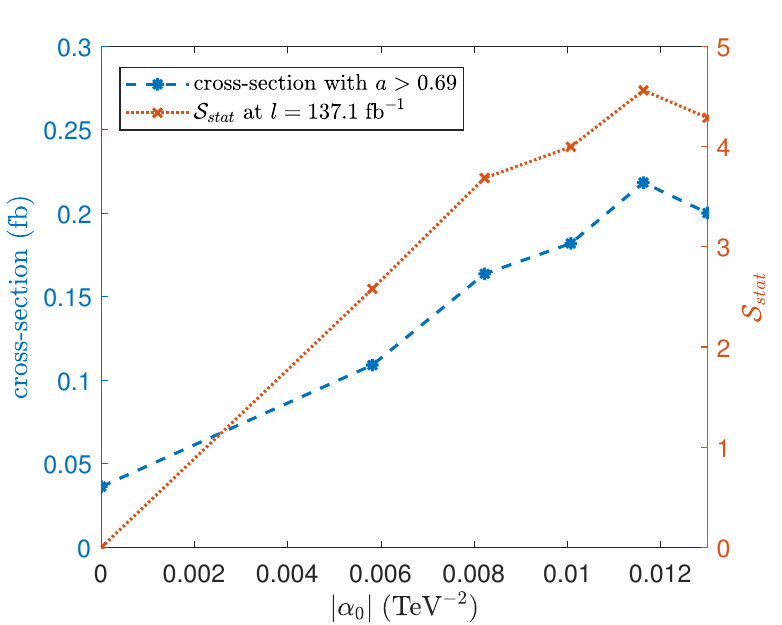}
\includegraphics[width=0.48\textwidth]{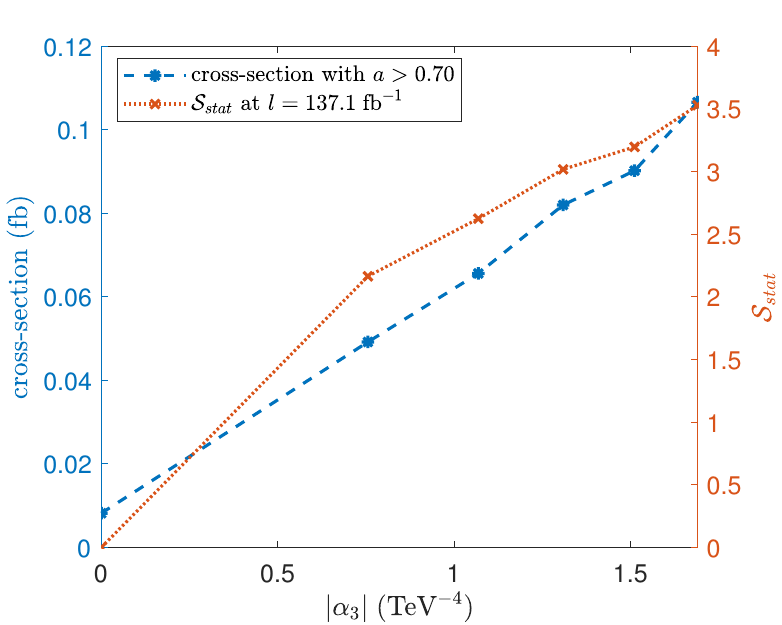}
\caption{\label{fig:cs2}The cross sections and signal significances after the anomaly score cut as functions of $\alpha _{0,3}$.}}
\end{figure}

Since the anomaly scores for the backgrounds increase a little, we use $a_{\rm min}=0.68$ for $V_0$ as a cut, and $a_{\rm min}=0.70$ for $V_3$.
The cross sections after this cut are shown in Fig.~\ref{fig:cs2}.
Using the cross sections, one can obtain the signal significance, which is $\mathcal{S}_{stat}=\sqrt{l} \times \left(\sigma (\alpha _i)-\sigma (\alpha _i=0)\right) / \sqrt{\sigma (\alpha _i)}$, where $\sigma (\alpha _i)$ is the cross section at $\alpha _i$ after the anomaly score cut.
The integrated luminosity currently at the LHC at $\sqrt{s}=13\;{\rm TeV}$ is about $l=137.1\;{\rm fb}^{-1}$~\cite{lumin}.
The signal significances at $l=137.1\;{\rm fb}^{-1}$ are also shown in Fig.~\ref{fig:cs2}.
If $|\alpha _0|=0.0058$, the signal of $V_0$ should be observed with $\mathcal{S}_{stat}=2.58$, therefore if the signal is not observed with $\mathcal{S}_{stat}>2$, the constraint is $|\alpha _0|<0.0058\;{\rm TeV}^{-2}$.
Similarly, for $\alpha_3$ the expected constraint is $|\alpha_3|<1.07\;{\rm TeV}^{-4}$.
The result of IF algorithm is better compared with the expected constraint at $\mathcal{S}_{stat}>2$ in Ref.~\cite{aqgc3} which is $\alpha _0\in [-0.0071, 0.0069]\;{\rm TeV}^{-2}$ and $\alpha _3\in [-1.73, 1.30]\;{\rm TeV}^{-4}$.
Again, for the cases that signal events are fewer, the IF algorithm performs better.

\section{\label{level4}Summary}

As more and more data are collected on the colliders, it becomes increasingly important to simplify the search of NP signals.
In this paper, we investigate a model independent approach for searching the NP signals which exploits the characteristics of the NP signals: few and kinematically different.
We use an unsupervised ML algorithm, also known as the IF algorithm, to find out the kinematically unusual events directly.

The IF algorithm is transparent and easy to apply.
This approach has the advantage that the suspected signals of the NP can be picked out without the knowledge of the NP models.
It works as an automatic ESS which can be generally applied.
We also show that the IF algorithm can be applied to constraint the parameters of NP models and coefficients of the operators.
Apart from that, the kinematically unusual events picked out are always worth studying.
There are also some limitations in this approach.
When anomalies appear, one needs to look deeper into them to know where they are originated.
Beyond that, there is room for improvement in how the data is organized.
In this paper, we directly use the components of the 4-momenta of the particles in the final state.

We use the dimension-8 operators contributing to the aQGCs as examples to investigate the capabilities of this approach.
The process $pp\to jj \ell^+\ell^-\nu\bar{\nu}$ is chosen as an arena, which has some complexity due to the neutrinos in the final state.
It can be shown that the anomaly scores of the background events are generally smaller than those of the signal events.
With a minimal allowed anomaly score as a cut, the signal events can be selected efficiently.
The IF algorithm shows greater ability to highlight the signal events and constraint the coefficients of the operators compared with the ESS designed for the aQGCs in this process.
In addition, we also show that IF algorithm performs better with fewer signal events.
The IF algorithm or other machine learning methods can be a very promising tool in the future study of high energy physics.

%\section*{ACKNOWLEDGMENT}

%\noindent
\begin{acknowledgments}
This work was supported in part by the National Natural Science Foundation of China under Grants No.11905093 and No.12047570, the Natural Science Foundation of the Liaoning Scientific Committee No.2019-BS-154 and the Outstanding Research Cultivation Program of Liaoning Normal University (No.21GDL004).
\end{acknowledgments}

\bibliography{IsolationForest}
\bibliographystyle{h-physrev}

\end{document}